# Sb-5s lone pair dynamics and collinear magnetic ordering in $Ba_2FeSbSe_5$

Stefan Maier [1,§,*], Michael W. Gaultois[2], Nami Matsubara[1,$], Wesley Surta[3], Francoise Damay[4], Sylvie Hebert[1], Vincent Hardy[1], David Berthebaud[5], Franck Gascoin[1,*]

1 Laboratoire CRISMAT UMR 6508 CNRS ENSICAEN, 6 boulevard du Maréchal Juin, 14050 Caen Cedex 04, France

2 Leverhulme Research Centre for Functional Materials Design, The Materials Innovation Factory, Department of Chemistry, University of Liverpool, 51 Oxford Street, Liverpool L7 3NY, United Kingdom

3 University of Liverpool, Department of Chemistry, Crown Street, Liverpool L69 7ZD, United Kingdom

4 Laboratoire Léon Brillouin, CEA, Centre National de la Recherche Scientifique, CE-Saclay, 91191 Gif-sur-Yvette, France

5 CNRS-Saint Gobain-NIMS, UMI 3629, Laboratory for Innovative Key Materials and Structures (LINK), National Institute for Materials Science, Tsukuba 305-0044, Japan

§ Current affiliation: Institute of Physics IA, RWTH Aachen University, 52074 Aachen, Germany

$ Current affiliation: KTH Royal Institute of Technology, SE-100 44, Stockholm, Sweden

* corresponding author: s.maier@physik.rwth-aachen.de

* corresponding author: franck.gascoin@ensicaen.fr

**Abstract**

Neutron diffraction and X-ray pair distribution function (XPDF) experiments were performed in order to investigate the magnetic and local crystal structures of $Ba_2FeSbSe_5$ and to compare them to the average (i.e. long-range) structural model, previously obtained by single crystal X-ray diffraction. Changes in the local crystal structure (i.e. in the second coordination sphere) are observed upon cooling from 295 K to 95 K resulting in deviations from the average (i.e. long-range) crystal structure. This work demonstrates, that these observations cannot be explained by local or long-range magnetoelastic effects involving Fe-Fe correlations. Instead, we found, that the observed differences between local and average crystal structure can be explained by Sb-5s lone pair dynamics. We also find, that below the Néel temperature ($T_N$ = 58 K), the two distinct magnetic $Fe^{3+}$ sites order collinearly, such that a combination of antiparallel and parallel spin arrangements along the *b*-axis results. The nearest-neighbor arrangement ($J_1$ = 6 Å) is fully antiferromagnetic, while next-nearest-neighbor interactions are ferromagnetic in nature.

**Introduction**

Wang et al. discovered $Ba_2FeSbSe_5$ in 2015. **[1]** Further studies of this compound then led to the discovery of rapid phase changes between an amorphous and a crystalline phase upon laser irradiation **[2]**, which is of significant interest for phase-change applications. It has been proposed, that the structural distortions around the antimony atoms due to stereoactive Sb-5s lone pairs play an important role in understanding the interaction of the sample surface with pulsed laser light. These findings sparked interest in studying these structural distortions in more detail. In this context the present study focuses on two aspects: magnetoelastic effects and Sb-5s lone pair stereoactivity. **Fig. 1** shows the crystal structure of $Ba_2FeSbSe_5$ including the connectivity of the $SbSe_6$ and $FeSe_4$ polyhedrons as well as the stereoactivity of the Sb-5s lone pairs in **Fig. 1c** and **Fig. 1d**. A second aspect of this study focuses on the magnetic structure of $Ba_2FeSbSe_5$ in order to rule out magnetoelastic effects involving changes in the Fe-Fe distances at 95 K and to lay the foundation for a better understanding of the previously observed phonon-mediated coupling between magnetic spins and electric dipoles at the Néel temperature (58 K) **[3]**, which is of significant interest for magnetodielectric technologies such as spin-charge transducers. The potential of $Ba_2FeSbSe_5$ as a magnetodielectric material, the diverse connectivity of the $FeSe_4$ tetrahedrons and $SbSe_6$ octahedrons (*cf.* **Fig. 1**) and different Fe-Fe distances also promise interesting arrangements and a complex magnetic coupling of the iron spins. Hence, neutron diffraction and X-ray pair distribution function (XPDF) experiments were carried out in order to gain a deeper understanding of the structural distortions in $Ba_2FeSbSe_5$, which helps understanding the observed phase-change and magnetodielectric properties of this remarkable semiconductor.

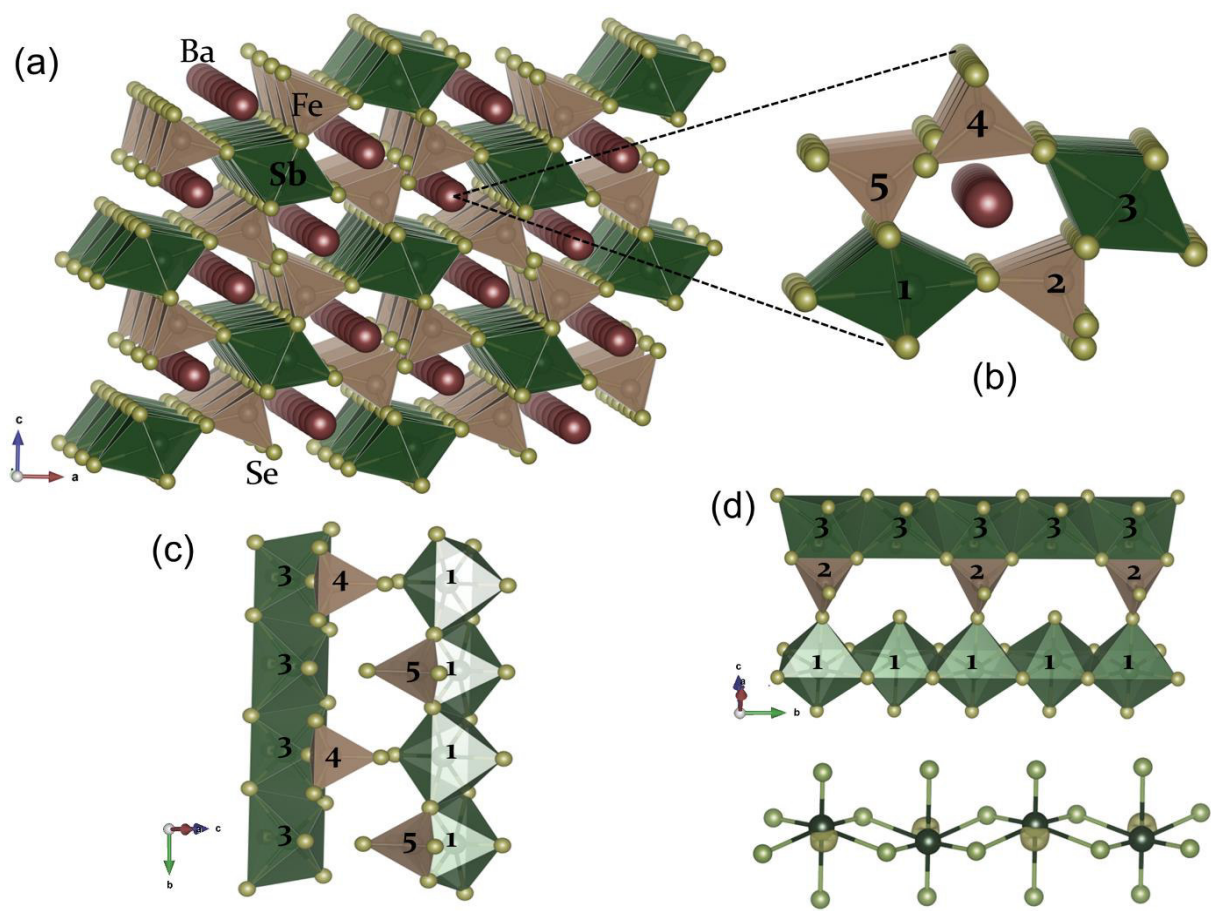

**Fig. 1.** (a) Crystal structure of $Ba_2FeSbSe_5$ (the unit cell is shown in **Fig. S1**); (b) local Ba environment, (c) and (d) show differently connected chain-like fragments within the Fe-Sb-Se network; (d) also shows the local distortions of the $SbSe_6$ octahedrons due to stereoactive lone pairs. Sb: green, Fe: brown, Se: light green, Ba: red.

**Experiments**

*Heat capacity measurements*

Heat capacity measurements were carried out in a Physical Properties Measurements System (PPMS, Quantum Design), by using a relaxation method with a 2τ analysis.[4]

*X-ray Pair Distribution Function (XPDF): experiments and data analysis*

All experiments were conducted at the I15-1 XPDF beamline at the Diamond Light Source (Didcot, UK) using a focused X-ray beam (20 μm in size) with a wavelength of 0.16 Å. Two dimensional X-ray scattering datasets were recorded up to a momentum transfer of 36 Å$^{-1}$. Densely packed, 0.7 mm borosilicate capillaries were prepared prior to data collection. The capillaries were placed horizontally on a 10 Hz capillary spinner and all samples were constantly rotated during data collection. The beamline is equipped with a Bent-Laue monochromator (700 μm horizontal focusing), a multi-layer mirror (20 μm vertical focusing) and a Perkin Elmer XRD1611 CP3 area detector with an active area of 409.6 × 409.6 mm². Data was collected at 295 K and at 95 K in order to increase the resolution by minimizing thermal vibrations. Low-temperature experiments were performed using a Cryojet operating with liquid nitrogen. The DAWN [5, 6] software was used for data processing and PDFgetX$_3$ [7] was used to convert 1D X-ray powder diffraction data to atomic PDFs. PDFfit2 was used to fit the experimental PDFs and PDFgui was used as a graphical user interface. [8] LaB$_6$ was used to extract the $Q_{damp}$ and $Q_{broad}$ parameters used for all subsequent refinements. All refinements were performed refining the scale factor, lattice parameters, delta1 (accounts for the correlated motion of atoms at low r, which sharpens the first peak(s) at low r), thermal displacement parameters and atomic positions. The thermal displacement parameters were constrained in a way so that $U_{11} = U_{22} = U_{33}$ and all U of the same atom type were assumed to be identical. Simultaneous fits of the PDF data and the reduced structure function ($F(Q)$) using large box models was done using the reverse Monte Carlo (RMC) method implemented in the RMCProfile software package. [9] Starting configurations were generated using Rietveld refinements from this study at the corresponding temperatures. Supercells with lattice parameters of 7*a* x 7*b* x 7*c* consisting of 12348 atoms were generated and fit using distance window constraints generated by analyzing structures from Rietveld refinements. Data were fit using a 1% tolerance on distance windows for interatomic distances for 24 h, followed by a 50% tolerance for 36 h. Simulations were run for 24 h with all distance windows for all pair correlations were removed except for Fe-Se (1.15 - 2.55 Å) and Sb-Se (2.45 - 9 Å), allowing them to fit the full breadth of low-*r* pair correlations but preventing over fitting. Finally, data were fit within five runs (*cf.* **Table S2** in the supplementary information) with no constraints or restraints to ensure persistence of the final configuration.

*Neutron powder diffraction*

Temperature dependent Neutron Powder Diffraction (NPD) experiments were performed on the G4.1 beamline (CEA-Saclay, France, λ = 2.426 Å) within a temperature range of 1.5 K - 300 K. Symmetry analysis and Rietveld refinements were performed using the FullProf Suite. [10] The former was carried out using the Bilbao Crystallographic Server. [11,12]

**Results and discussion**

Signature of Sb-5s lone pair dynamics in the X-ray pair distribution function

X-ray pair distribution function (XPDF) data were recorded at the I15-1 XPDF beamline at the Diamond Light Source (Didcot, UK). Two datasets were recorded at 295 K and 95 K (which is the lowest reachable temperature with this experimental setup) in order to increase the resolution by minimizing thermal vibrations. **Fig. 2** shows the experimental pair distribution functions (PDFs) with the corresponding fits based on the structural model (space group: *Pnma*) of the average (i.e. long-range) crystal structure obtained from single crystal diffraction experiments performed by Wang et al. [1]

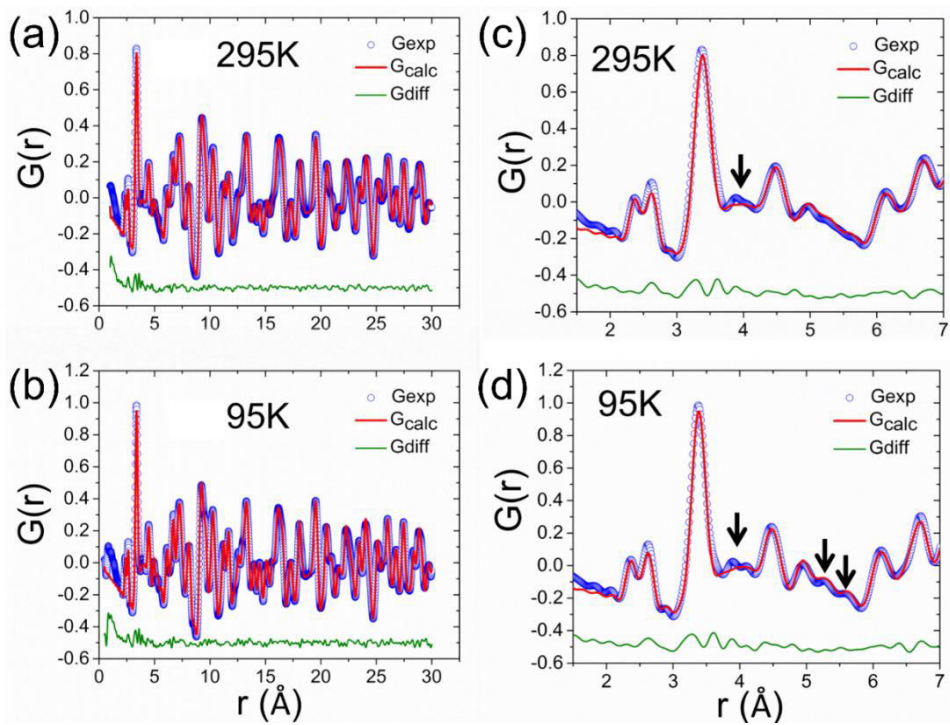

**Fig. 2:** Calculated (red) and experimental (blue) pair distribution functions (PDFs) with the corresponding difference plot (green) of crystalline $Ba_2FeSbSe_5$ at (a) 295 K and (b) 95 K; (c) and (d) show the low r range of the PDFs illustrating the changes in the local crystal structure upon cooling (*cf.* black arrows). All fits are based on a structural model, which represents the average (i.e. long-range) atomic structure previously obtained from single crystal X-ray diffraction experiments (*cf.* Ref. [1]).

In general, the calculated and experimental PDFs agree quite well. A closer inspection of the low r-range (i.e. from 2-7 Å), however, revealed clear misfits and changes in the atomic correlations between 3.8 Å and 5.6 Å upon cooling (*cf.* **Fig. 2c and d**). More specifically, we observe a peak splitting at approximately 4 Å and the sharpening of features at 5.3 Å and 5.6 Å. These features are not well described by the small box model (space group: Pnma) based on the average structural model obtained by Wang et al. With this publication we demonstrate, that these observations are signatures of Sb-5s lone pair dynamics. A detailed discussion in the supplementary information will help convince the critical reader, that these observations are not a result of measurement artifacts such as the termination of the diffraction data at finite momentum transfer (i.e. $Q_{max} = 36$ Å$^{-1}$). It has previously been shown [2], that a stereoactive lone-pair electron density is localized at the Sb atoms, which results in a distorted octahedral arrangement of the Se atoms surrounding Sb (*cf.* **Fig. 1d**). In the following we will demonstrate that the peak splitting at r ≈ 4 Å (*cf.* **Fig. 1d**) is a clear signature of this Sb-5s lone pair effect involving Sb-Se and Se-Se correlations. At 295 K, these are not as pronounced. There can be two reasons for this: a) the resolution is not high enough to resolve the subtle differences in the Sb-Se and Se-Se correlations at 295 K due to thermal broadening or b) the SbSe$_6$ locally appear undistorted (i.e. the Sb-5s lone pairs are rather spherical) due to thermal vibrations of the atoms and can therefore not be resolved. In the following we will demonstrate, that the differences we observe in our PDF data between local and average (i.e. long-range) crystal structure involve Sb-5s lone pair dynamics. In order to remove any ambiguity from the interpretation of our observations it is required to rule out any magnetoelastic effects related to the iron spins, which is done in the next section.

### Ruling out magnetoelastic effects due to local magnetic pre-ordering of the Fe spins

We previously found a coupling between phonons, magnetic spins and electric dipoles at the Néel temperature (*cf.* Ref [3]), which is equivalent to magnetoelastic effects since electric dipoles can only form, when the atoms are moving out of their equilibrium position, which happens at the same time as the magnetic spins are ordering. It is therefore necessary to rule out out any magnetoelastic effects (i.e. changes in Fe-Fe distances due to short-range magnetic pre-ordering of the iron spins) at 95 K. It is important to keep in mind, that macroscopic properties such as heat capacity ($c_p$) and magnetic susceptibility ($\chi$) will only be influenced significantly by long-range magnetoelastic effects. This can be seen from **Fig. 3**, which shows the heat capacity ($c_p$) and magnetic susceptibility ($\chi$) as a function of temperature (a ZFC vs. FC comparison for $\chi(T)$ can be found in the supplementary information). The $\chi(T)$ and $c_p(T)$ curves show only one transition at the Néel temperature ($T_N$= 58 K), which is consistent with previous reports on the magnetic properties of Ba$_2$FeSbSe$_5$. **[1,3]**

Hence, from a macroscopic point of view, there are no long-range magnetoelastic effects occurring between 295K and 95 K, which would cause changes in χ and $c_p$ due to long-range magnetic ordering.

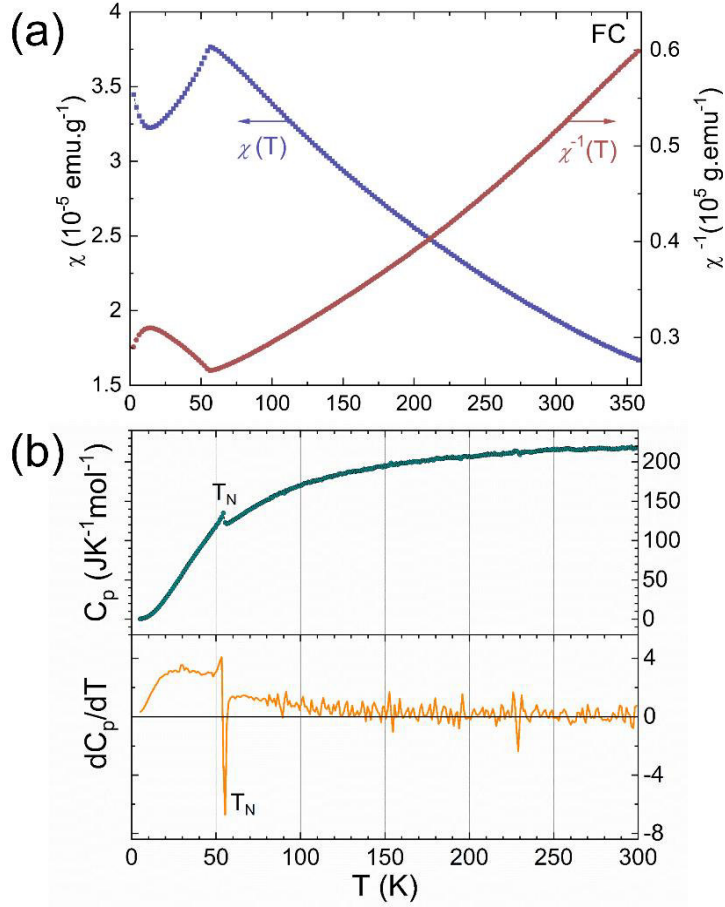

**Fig. 3.** (a) Magnetic susceptibility χ(T) of $Ba_2FeSbSe_5$ as a function of temperature and (b) top channel: heat capacity ($c_p$) and bottom channel: $dc_p/dT$ showing only one transition corresponding to long-range magnetic ordering in crystalline $Ba_2FeSbSe_5$ below $T_N$. Hence, no long-range magnetoelastic effects can be observed between 300 K and 95 K. The magnetic moment obtained from (a) is 6.8 $μ_B$.

There is, however the possibility of local magnetic pre-ordering (accompanied by a slight change in the Fe-Fe distances) or Se-mediated magnetic super exchange between the Fe atoms. Both scenarios do not result in long-range magnetic ordering and could be missed by only looking at macroscopic properties such as magnetic susceptibility and the heat capacity. To rule out local magnetoelastic effects involving changes in the Fe-Fe distances at 95 K, we compared structural models obtained from neutron diffraction and PDF data. It is important to note, that we compare the average (i.e. long-range) crystal structure (i.e. only the nuclear part in the fully ordered state) obtained from neutron diffraction and PDF data. Before doing so, it is necessary to obtain a model of the long-range ordered nuclear and magnetic structure from neutron diffraction data.

Neutron powder diffraction (NPD) experiments were carried out between 1.5K and 300K. **Fig.4a** shows the temperature evolution of the NPD patterns, which confirms the long-range magnetic ordering below 58K, manifesting itself in the appearance of additional Bragg reflections. The temperature evolution of the lattice parameters can be found in **Fig. S3** in the supplementary information.

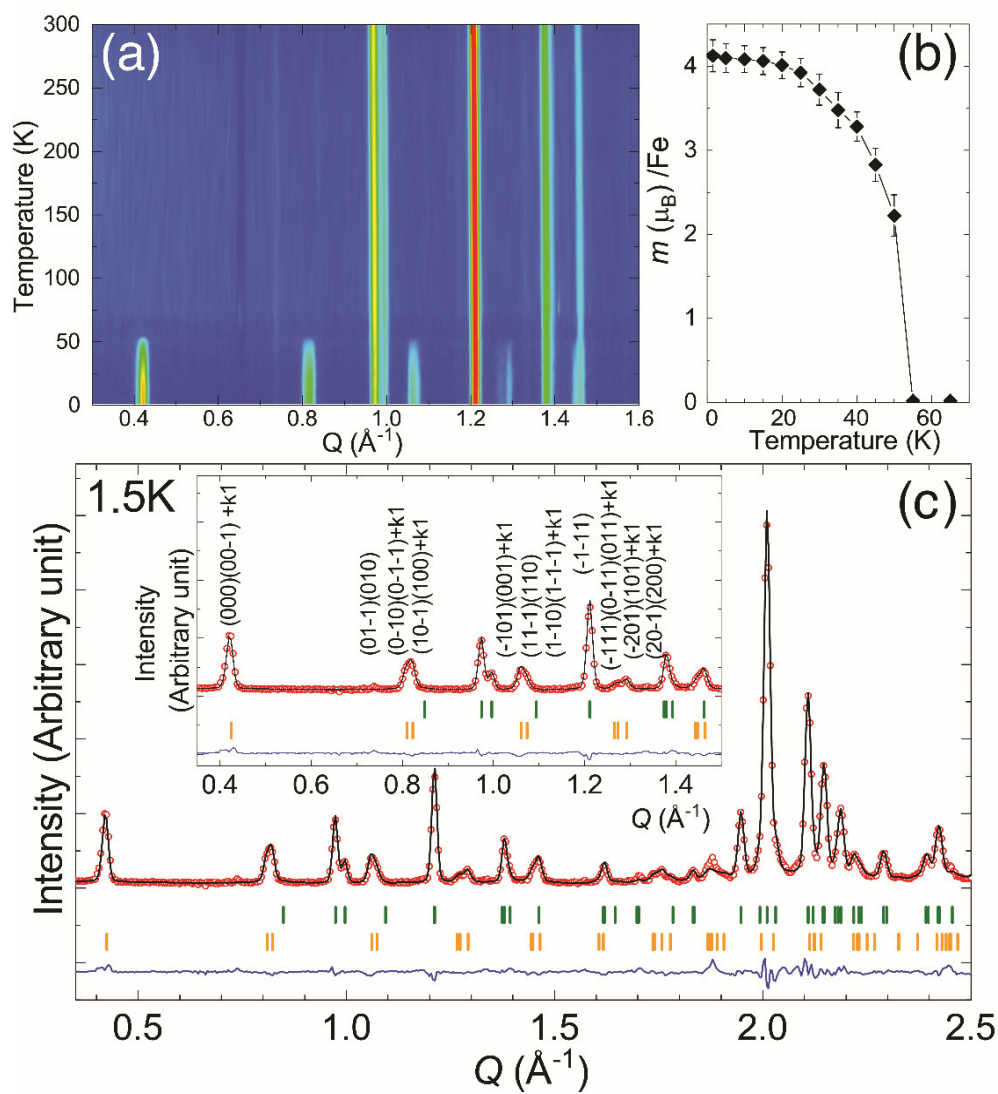

**Fig. 4:** (a) Neutron powder diffraction (NPD) patterns as a function of temperature showing magnetic reflections appearing below $T_N$; (b) magnetic moment (constrained) of Fe as a function of temperature; (c) Rietveld refinement ($R_{wp}$ = 4.58 %, background corrected $R_{wp}$ = 9.93 %, magnetic R = 7.59 %) of the diffraction pattern recorded at 1.5K including the crystal and magnetic model: experimental data: open red circles; calculated profile: black continuous line; allowed Bragg reflections of the crystal (i.e. nuclear) and magnetic structure: green and orange vertical marks, respectively. The difference between the experimental and calculated profiles is displayed at the bottom of the graph as a blue continuous line. Please note, that the neutron diffraction profiles of the crystal structure are based on a structural model describing the average (i.e. long-range) atomic arrangement, which makes them comparable to the PDF model (*cf.* **Fig.5**).

All the magnetic Bragg reflections can be indexed with the commensurate propagation vector **k** = (½ 0 ½) as can be seen from **Fig. 4c**. In order to constrain the number of solutions for the magnetic model, a symmetry analysis was carried out for the Fe Wyckoff site 4e. There are eight irreducible representations with one dimension, each containing either 1 or 2 basis vectors. The best agreement between experimental data and our model is obtained for the $\Gamma_4$ and $\Gamma_8$ representations, whose basis vectors are summarized in Table 1.

**Table 1**: Irreducible representations ($\Gamma_4$ and $\Gamma_8$) for the 4e Fe site in the space group *Pnma* with **k** = (1/2, 0, 1/2).

| Irreducible representations | | x,y,z | -x+1/2,-y+1,z+1/2 | -x+1,y+1/2,-z+1 | x+1/2,-y+1/2,-z+1/2 |
|---|---|---|---|---|---|
| $\Gamma_4$ | Re | 0 1 0 | 0 0 0 | 0 -1 0 | 0 0 0 |
|  | Im | 0 0 0 | 0 1 0 | 0 0 0 | 0 -1 0 |
| $\Gamma_8$ | Re | 0 1 0 | 0 0 0 | 0 -1 0 | 0 0 0 |
|  | Im | 0 0 0 | 0 -1 0 | 0 0 0 | 0 1 0 |

Both irreducible representations result in the Shubnikov group $P_a2_1/m$ (#11.55), which has two independent, magnetic Fe sites within this magnetic symmetry. Only one magnetic component $m_y$ can be refined independently. However, a magnetic model where only one out of two Fe sites is magnetically ordered results in a magnetic moment that is slightly too high (~ 5.8 $\mu_B$) for an $Fe^{3+}$ state with S = 5/2 (expected 5 $\mu_B$) and is unphysical. Thus, the magnetic moment of both iron sites was constrained to be the same since they are crystallographically identical. For data recorded at 1.5 K, the Rietveld refinement yields $m_y$ = 4.13 (4) $\mu_B$, which is close to the expected value of 5 $\mu_B$ for an $Fe^{3+}$ state with S = 5/2 and to the experimental moment of 5.3 $\mu_B$ found by Wang et al. [1] The temperature evolution of the refined magnetic moment is depicted in **Fig. 4b**. The magnetic moment obtained from our magnetic susceptibility data (*cf.* **Fig. 3** and **Fig. S4**) is slightly higher (i.e. 6.8 $\mu_B$). A more detailed discussion concerning this point can be found in the supplementary information. The model we obtain is a collinear magnetic structure with magnetic moments pointing along the *b*-axis of the crystal structure (*cf.* **Fig. 5**). The magnetic structure can be visualized as two distorted triangular $Fe^{3+}$ ladders running along the *a*- and *b*-axis (*cf.* **Fig. 5a**) with Fe-Fe distances ranging from 6.0 Å to 9.1 Å. Within the triangular ladders along the *b*-axis, the nearest-neighbour arrangement ($J_1$ at a distance of 6.0 Å) can be understood as a zig-zag antiferromagnetic pattern (*cf.* grey, dotted line in **Fig. 5a**) whereas $J_3$ and $J_4$ correspond to ferromagnetic interactions (*cf.* thin green line in **Fig. 5a**). The ladders running along the *a*-axis can be described as two antiferromagnetic chains ($J_2$ at distance of 6.5 Å), which have an up-up-down-down spin structure resulting in geometrical frustration.

Ba$_2$FeSbSe$_5$ shows antiferromagnetic ordering ($\Theta$ = -133 (2) K) and a rather high transition temperature (T$_N$ = 58K), despite the rather large Fe-Fe distances (> 6.0 Å). Similar spin arrangements were observed in antiperovskite chalco-halides such as Ba$_3$(FeS$_4$)Br, which also consists of FeS$_4^{5-}$ tetrahedrons. [13] In this case, the shortest Fe-Fe distances are 6.2 Å, i.e. similar to those found in Ba$_2$FeSbSe$_5$. And yet, a clear antiferromagnetic spin arrangement has also been observed for Ba$_3$(FeS$_4$)Br. The magnetic interaction between the Fe atoms was rationalized by a Fe-S-S-Fe super exchange with a strong hybridization between the Fe $d$ orbitals and the S $p$ orbitals. The $J_1$ interaction in Ba$_2$FeSbSe$_5$ can be explained in an analog way by a Fe-Se-Se-Fe super exchange, which will be discussed further in the next section.

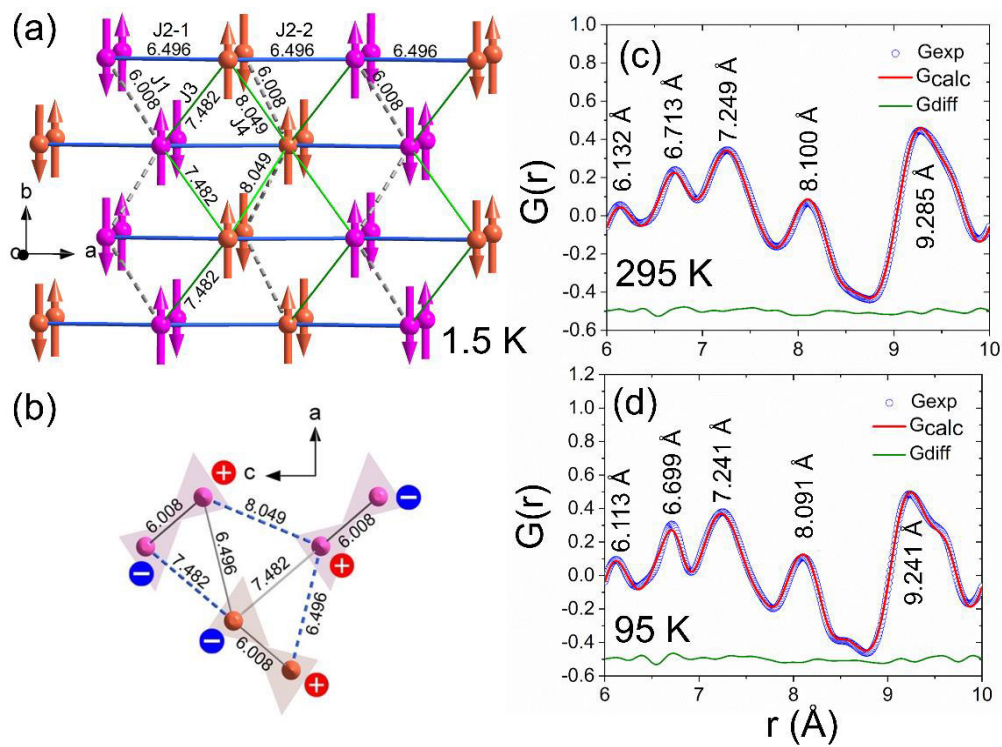

**Fig. 5** Projections of the magnetic structure of Ba2FeSbSe5 in the (a) a-b plane and (b) in the a-c plane based on the refinement shown in **Fig.4**. Two independent, magnetic Fe$^{3+}$ sites - orange (Fe1) and pink (Fe2) - are distinguished in each projection. Fe-Fe distances in the triangular ladders are indicated; (c)-(d) Calculated (red) and experimental (blue) pair distribution functions (PDFs) showing the r-range of the Fe-Fe correlations at 295 K and 95 K, respectively. No misfits or changes in these correlations is visible, i.e. there are no differences between this part of the local and average crystal structure unlike those found for the Se-Se correlations depicted in **Fig. 6**. Please note, that both, the calculated PDF and neutron diffraction profiles of the atomic structure are based on a structural model describing the average (i.e. long-range) atomic arrangement, which makes them comparable. The interatomic distances shown in (a) and (b) were extracted from our NPD model obtained at 1.5 K (*cf.* **Fig. 4** and **Table S1**), while those depicted in (b) correspond to the PDF peak positions (i.e. the pairwise correlations). Slight differences in these interatomic distances are natural due to thermal expansion (*cf.* **Table S1**).

After establishing the model for the magnetic structure of $Ba_2FeSbSe_5$ at 1.5 K, we now compare the nuclear part to the local crystal structure obtained from our PDF data. The main focus lies on the Fe-Fe correlations in order to draw conclusions with the respect to a local magnetoelastic coupling within the triangular framework involving a displacement of the Fe atoms as it has been reported for $AgCrS_2$, $CuMnO_2$, $CuCrO_2$ and $NaMnO_2$ [14-16]. From **Table S1** in the supplementary information and from **Fig. 5 c and d**, it is clear, that the local Fe-Fe correlations do not change upon cooling (slight changes due to thermal expansion are natural) and the average structural model describes these correlations well. Hence, both local and long-range magnetoelastic effects connected to changes in the Fe-Fe correlations cannot be observed by the experimental methods employed here. They can therefore be ruled out as being the origin of the observed differences between local and average (i.e. long-range) crystal structure in the PDF data. Se-mediated Fe-Se-Se-Fe super exchange between the Fe atoms, however, cannot be ruled out, which is discussed in the next section.

<u>Sb-5s lone pair dynamics vs. Se-mediated magnetic super exchange</u>

There is a series of studies, which contributed to the understanding of local cation off-centering in inorganic or hybrid organic-inorganic solids. **[17-27]** All these publications demonstrated how powerful XPDF analyses and experiments are when investigating differences between average (i.e. long-range) and local crystal structure. In this section, we discuss the observed peak splitting at approximately 4 Å and the sharpening of PDF features at 5.3 Å and 5.6 Å upon cooling in the context of two phenomena, which are difficult (if not impossible) to disentangle in the case of $Ba_2FeSbSe_5$: Sb-5s lone pair dynamics and Se-mediated magnetic super exchange. **Fig. 6 c and d** show the experimental and calculated pair distribution functions in the r-range of the Sb-Se and Se-Se correlations (*cf.* **Fig. 7**), which surround the asymmetrically coordinated antimony atoms (*cf.* **Fig. 6a**). It is important to note, that the $SbSe_6$ octahedrons are connected to the $FeSe_4$ tetrahedrons via edges and corners (*cf.* **Fig. 6 a and b**), i.e. Se-mediated magnetic super exchange as well as lone pair dynamics have to be considered as possible origins of the observed changes in the local crystal structure of $Ba_2FeSbSe_5$. However, no experimental evidence of such effects was found, which would be visible as subtle changes in the Fe-Fe correlations resulting in differences between local and average crystal structure.

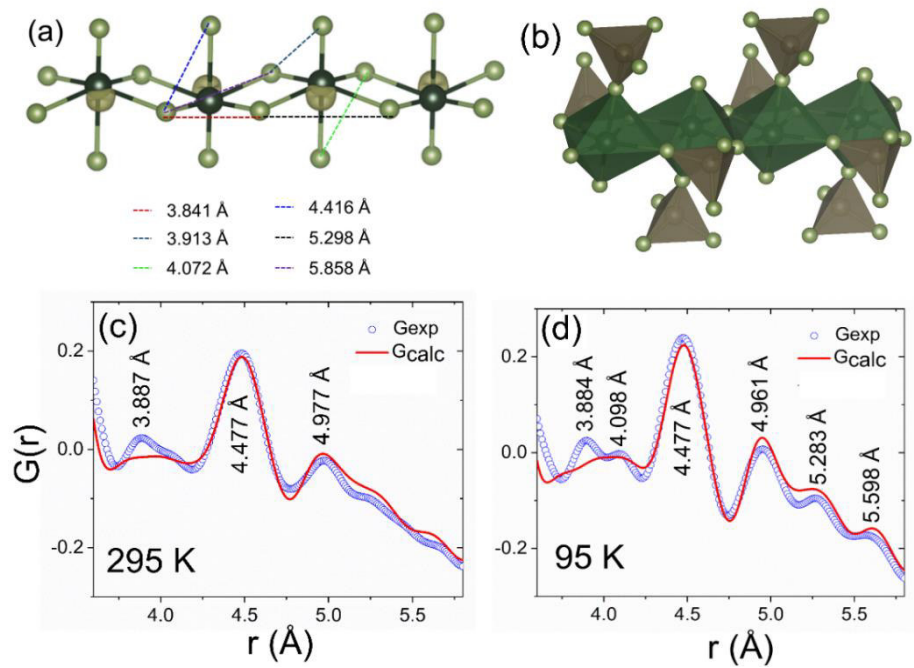

**Fig. 6** (a) Local distortions of the SbSe$_6$ octahedrons due to stereoactive lone pairs (yellow) with the corresponding distances from the average crystal structure obtained by Wang et al. [1] (b) connectivity of the SbSe$_6$ octahedrons and the FeSe$_4$ tetrahedrons. Calculated (red) and experimental (blue) pair distribution functions (PDFs) in the r-range of the Se-Se correlations at (c) 300 K and (d) 95 K. Misfits between the calculated (i.e. average crystal structure) and the experimental PDFs (i.e. local crystal structure) as well as changes in the local crystal structure upon cooling are clearly visible. Please note, that the interatomic distances depicted in (c) and (d) correspond to the peak position of the pairwise correlations obtained from the experimental PDF data. Slight differences between these interatomic distances and the ones shown in (a) are due to thermal expansion and the use of different experimental methods. However, the accuracy of all interatomic distances is sufficient to conclude, that Sb-5s lone pair dynamics cause the observed peak splitting at approximately 4 Å and the sharpening of the PDF features at 5.3 Å and 5.6 Å. A comparison of all relevant interatomic distances can be found in **Table S1** in the supplementary information.

Reverse Monte Carlo (RMC) simulations were performed in order to better describe the experimental PDF data in the area, where the local and average crystal structure differ and where changes in the PDFs with temperature are observed. **Fig 7a-b** show the improved model of the local crystal structure at 295 K and 95 K also describing the r-range of the Sb-Se and Se-Se pair correlations well (the model for the entire experimental r-range can be found in **Fig. S7**). They also include the respective contributions of the Sb-Se and Se-Se pair correlations, while **Fig. 7c and d** depict the corresponding changes in Sb-Se and Se-Se pair correlations with temperature, where thermal broadening effects (i.e. changes in anisotropic displacement parameters) have to be kept in mind. However, this figure provides strong evidence for Sb-5s lone pair dynamics being responsible for subtle differences between average and local crystal structure.

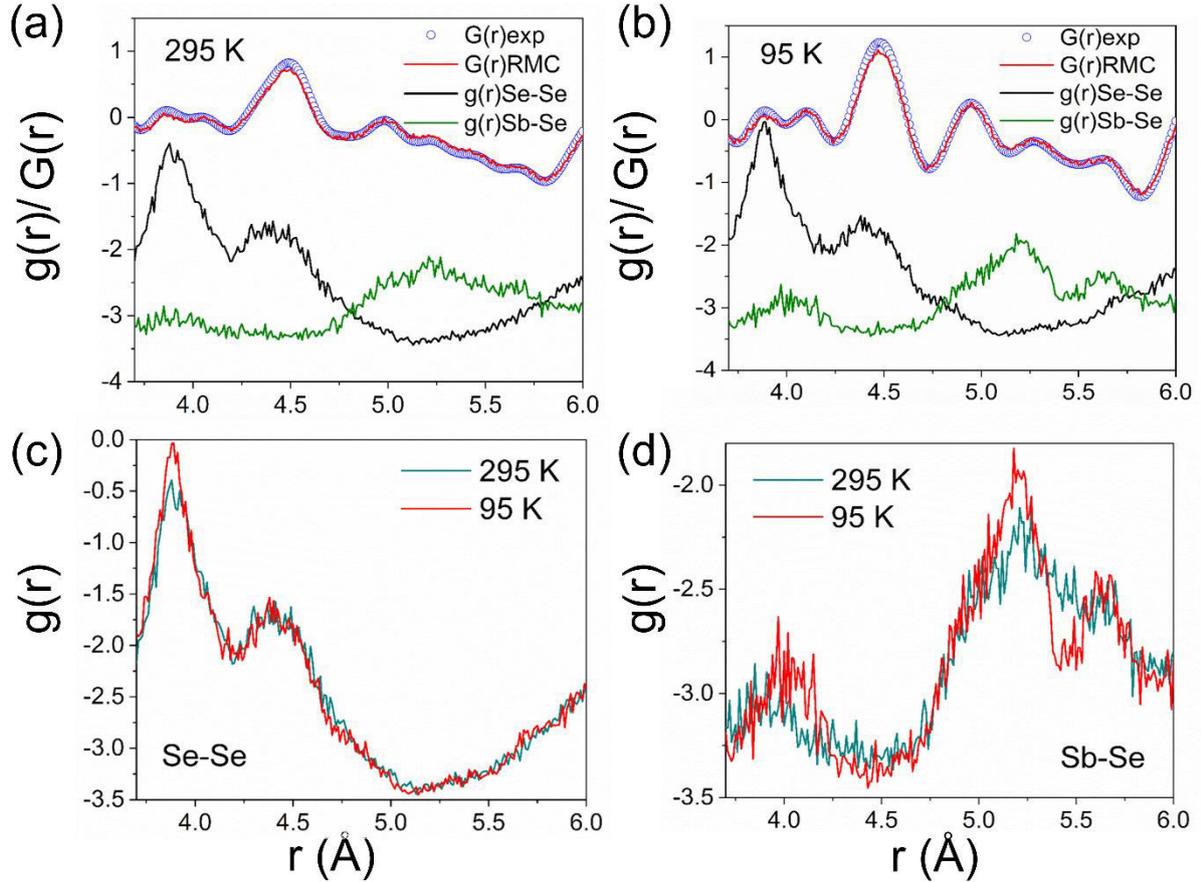

**Fig. 7** RMC fits of the experimental PDF data at (a) 295 K and (b) 95 K with the corresponding Sb-Se and Se-Se pair correlations. The RMC simulations result in a better description of the experimental data, which (due to Sb-5s lone pair effects) cannot be well described by the average structural model in the region of the Sb-Se and Se-Se correlations; (c) and (d) show subtle changes in the Sb-Se and Se-Se pair correlations with temperature. This figure provides evidence for Sb-5s lone pair dynamics being responsible for subtle differences between average and local crystal structure.

It is important to note, that thermal broadening is an effect affecting all pairwise correlations (e.g. Fe-Fe). However, the small box model based on the average structural model obtained by Wang et al. only fails to describe the r-range of the Sb-Se and Se-Se pair correlations adequately. This provides further evidence, that what we observe in the XPDF data is an experimental signature of Sb-5s lone pair dynamics locally changing the Sb-Se and Se-Se correlations with respect to the average (i.e. long-range) atomic arrangement. Hence, these findings strongly support the previous proposal, that the structural distortions around the antimony atoms due to dynamic stereoactive lone pairs may play an important role in understanding the interaction of the sample surface with pulsed laser light. [2] However, further in situ XPDF experiments during laser irradiation are required to fully understand the amorphization process of $Ba_2FeSbSe_5$, when interacting with pulsed laser light.

**Conclusion**

X-ray pair distribution function (XPDF) experiments revealed differences between average (i.e. long-range) and local crystal structure as well as changes in the local crystal structure upon cooling from 295 K to 95 K, which is well above $T_N$ = 58 K. This work demonstrates that these observations can be explained by Sb-5s lone pair dynamics locally changing the Sb-Se and Se-Se correlations and that these observations are not caused by local or long-range magnetoelastic effects involving Fe-Fe correlations. In order to rule out these effects, the crystal structure of the magnetically ordered phase was compared to the average crystal structure of the paramagnetic state (i.e. at 295K). In order to make this comparison, a model of the magnetic structure below $T_N$ (i.e. at 1.5 K) was obtained. The magnetic structure can be described as two distinct magnetic $Fe^{3+}$ sites ordering collinearly, such that a combination of antiparallel and parallel spin arrangements along the *b*-axis results. The nearest-neighbor arrangement ($J_1$ = 6 Å) is antiferromagnetic, while next-nearest-neighbor interactions are ferromagnetic.

**Acknowledgements**

The authors acknowledge the Diamond Light Source for time on the I15-1 Beamline under the proposal EE16536 and in particular the support of Michael Wharmby during the beamtime. Furthermore, we acknowledge SOLEIL for providing NPD beamtime. Financial support from the French "Agence Nationale de la Recherche" (ANR) through the program "Investissements d'Avenir" (ANR-10-LABX-09-01), LabEx EMC3 is gratefully acknowledged.

**References**

[1] J. Wang, J. T. Greenfield, K. Kovnir, *J. Solid State Chem.* **2016**, 242, 22.

[2] S. Maier, S. Hebert, H. Kabbour, D. Pelloquin, O. Perez, D. Berthebaud, F. Gascoin, *Mater. Chem. Phys.* **2018**, 203, 202.

[3] S. Maier, C. Moussa, D. Berthebaud, F. Gascoin, A. Maignan, *Appl. Phys. Lett.* **2018**, 112, 202903.

[4] J. C. Lashley, M. F. Hundley, A. Migliori, J. L. Sarrao, P. G. Pagliuso, T. W. Darling, M. Jaime, J. C. Cooley, W. L. Hults, L. Morales, D. J. Thoma, J. L. Smith, J. Boerio-Goates, B. F. Woodfield, G. R. Stewart, R. A. Fisher, N. E. Phillips, *Cryogenics* **2003**, 43, 369.

[5] M. Basham, J. Filik, M. T. Wharmby, P. C. Y. Chang, B. El Kassaby, M. Gerring, J. Aishima, K. Levik, B. C. A. Pulford, I. Sikharulidze, D. Sneddon, M. Webber, S. S. Dhesi, F. Maccherozzi, O. Svensson, S. Brockhauser, G. Náray, A. W. Ashton, *J. Synchrotron Rad.* **2015**, 22, 853.

[6] J. Filik, A. W. Ashton, P. C. Y. Chang, P. A. Chater, S. J. Day, M. Drakopoulous, M. W. Gerring, M. L. Hart, O. V. Magdysyuk, S. Michalik, A. Smith, C. C. Tang, N. J. Terrill, M. T. Wharmby, H. Wilhelm, *J. Appl. Cryst.* **2017**, 50, 959.

[7] P. Juhás, T. Davis, C. L. Farrow, S. J. L. Billinge, *J. Appl. Cryst.* **2013**, 46, 560.

[8] C. L. Farrow, P. Juhás, J. W. Liu, D. Bryndin, E. S. Božin, J. Bloch, Th. Proffen, S. J. L. Billinge, *J. Phys.: Condens. Mat.* **2007**, 19, 335219.

[9] M. G. Tucker, D. A. Keen, M. T. Dove, A. L. Goodwin, Q. Hui, *J. Phys.: Condens. Matter* **2007**, *19*, 335218.

[10] J. Rodríguez-Carvajal, *Phys. Rev. B* **1993**, 192, 55.

[11] M. I. Aroyo, J. M. Perez-Mato, C. Capillas, E. Kroumova, S. Ivantchev, G. Madariaga, A. Kirov, H. Wondratschek, *Z. Cryst.* **2006**, 221, 15.

[12] M. I. Aroyo, A. Kirov, C. Capillas, J. M. Perez-Mato, H. Wondratschek, *Acta. Cryst.* **2006**, A62, 115.

[13] X. Zhang, K. Liu, J.-Q. He, H. Wu, Q.-Z. Huang, J.-H. Lin, Z.-Y. Lu, F.-Q. Huang, *Sci. Rep.* **2015**, 5, 1.

[14] F. Damay, C. Martin, V. Hardy, G. André, S. Petit, A. Maignan, *Phys. Rev. B* **2011**, 83, 184413.

[15] A. V. Ushakov, S. V. Streltsov, D. I. Khomskii, *Phys. Rev. B* **2014**, 89, 024406.

[16] M. Poienar, F. Damay, C. Martin, J. Robert, S. Petit, *Phys. Rev. B* **2010**, 81, 104411.

[17] G. Laurita, D. H. Fabini, C. C. Stoumpos, M. G. Kanatzidis, R. Seshadri, *Chem. Sci.* **2017**, 8, 5628.

[18] M. Giot, L. C. Chapon, J. Androulakis, M. A. Green, P. G. Radaelli, A. Lappas, *Phys. Rev. Lett.* **2007**, 247211, 99, 247211.

[19] E. S. Božin, C. D. Malliakas, P. Souvatzis, T. Proffen, N. A. Spaldin, M. G. Kanatzidis, S. J. L. Billinge, *Science* **2010**, 330, 1660.

[20] B. Sangiorgio, E. S. Božin, C. D. Malliakas, M. Fechner, A. Simonov, M. G. Kanatzidis, S. J. L. Billinge, N. A. Spaldin, T. Weber, *Phys. Rev. Mater.* **2018**, 2, 085402.

[21] K. M. Ø. Jensen, E. S. Božin, C. D. Malliakas, M. B. Stone, M. D. Lumsden, M. G. Kanatzidis, S. M. Shapiro, S. J. L. Billinge, *Phys. Rev. B* **2012**, 86, 085313.

[22] K. R. Knox, E. S. Božin, C. D. Malliakas, M. G. Kanatzidis, S. J. L. Billinge, *Phys. Rev. B* **2014**, 89, 014102.

[23] R. Yu, E. S. Božin, M. Abeykoon, B. Sangiorgio, N. A. Spaldin, C. D. Malliakas, M. G. Kanatzidis, S. J. L. Billinge, *Phys. Rev. B* **2018**, 98, 144108.

[24] H. Takenaka, I. Grinberg, A. M. Rappe, *Phys. Rev. Lett.* **2013**, 110, 147602.

[25] D. Hou, C. Zhao, A. R. Paterson, S. Li, J. L. Jones, *J. Eur. Ceram. Soc.* **2018**, 971.

[26] G. Laurita, K. Page, S. Suzuki, R. Seshadri, *Phys. Rev. B* **2015**, 92, 214109.

[27] M. Dutta, K. Pal, U. V. Waghmare, K. Biswas, *Chem. Sci.* **2019**, 10, 4905.

# Supplementary information: Sb-5s lone pair dynamics and collinear magnetic ordering in Ba$_2$FeSbSe$_5$

Stefan Maier [1,§,*], Michael W. Gaultois[2], Nami Matsubara[1,$], Wesley Surta[3], Francoise Damay[4], Sylvie Hebert[1], Vincent Hardy[1], David Berthebaud[5], Franck Gascoin[1,*]

1 Laboratoire CRISMAT UMR 6508 CNRS ENSICAEN, 6 boulevard du Maréchal Juin, 14050 Caen Cedex 04, France

2 Leverhulme Research Centre for Functional Materials Design, The Materials Innovation Factory, Department of Chemistry, University of Liverpool, 51 Oxford Street, Liverpool L7 3NY, United Kingdom

3 University of Liverpool, Department of Chemistry, Crown Street, Liverpool L69 7ZD, United Kingdom

4 Laboratoire Léon Brillouin, CEA, Centre National de la Recherche Scientifique, CE-Saclay, 91191 Gif-sur-Yvette, France

5 CNRS-Saint Gobain-NIMS, UMI 3629, Laboratory for Innovative Key Materials and Structures (LINK), National Institute for Materials Science, Tsukuba 305-0044, Japan

§ Current affiliation: Institute of Physics IA, RWTH Aachen University, 52074 Aachen, Germany

$ Current affiliation: KTH Royal Institute of Technology, SE-100 44, Stockholm, Sweden

* corresponding author: s.maier@physik.rwth-aachen.de

* corresponding author: franck.gascoin@ensicaen.fr

Table of content:

1. <u>Crystal structure and unit cell</u>

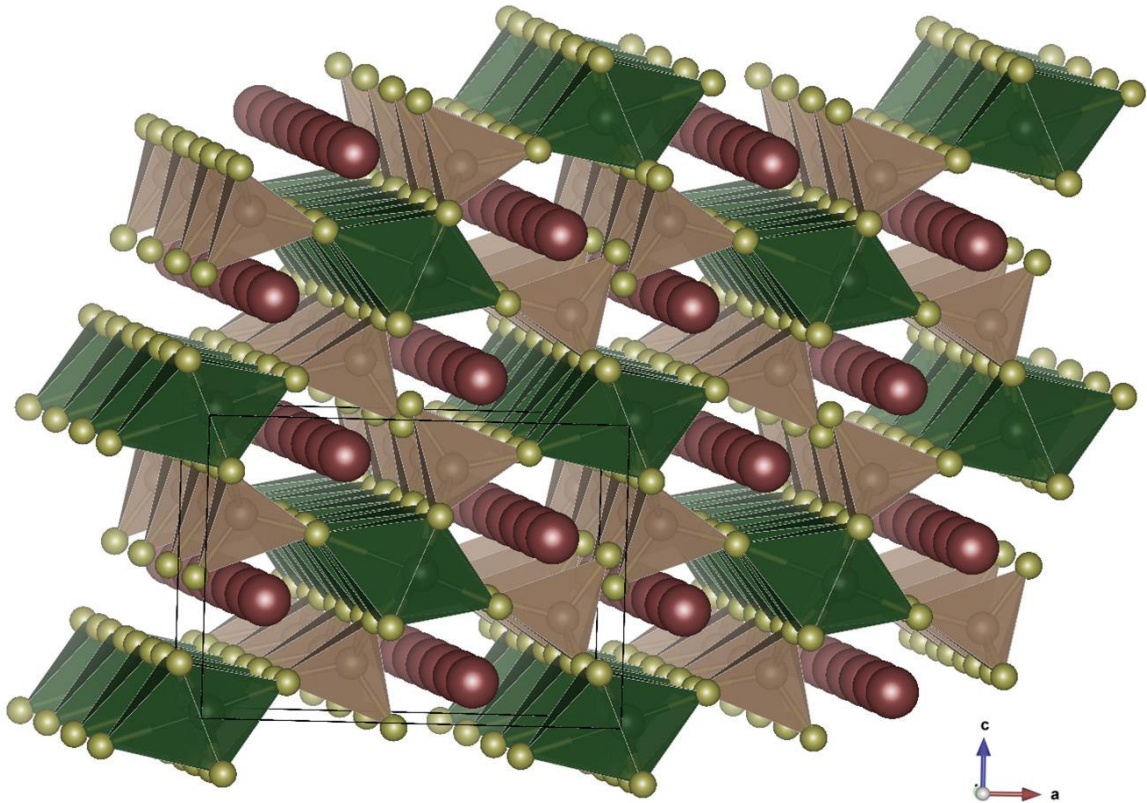

**Fig. S1:** $Ba_2FeSbSe_5$ crystal structure including the unit cell. Sb: green, Fe: brown, Se: light green, Ba: red.

2. <u>Ruling out measurement artifacts in the PDF data</u>

There are two effects, which need to be kept in mind when interpreting the presented PDF data. a) cooling from 295 K to 95 K will increase the resolution since thermal vibrations are reduced, which leads to a thermal broadening of the peaks (this is why we performed the cooling experiment) and b) terminating the diffraction data at a finite momentum transfer causes so-called termination ripples (see below) and decreases the resolution. **[S1]** In our case we always used the same cut-off of $Q_{max}$ = 36 Å$^{-1}$, i.e. the resolution is the same with respect to the Q cut-off and the termination ripples do therefore not differ either (*cf.* red boxes in **Fig. S2**).

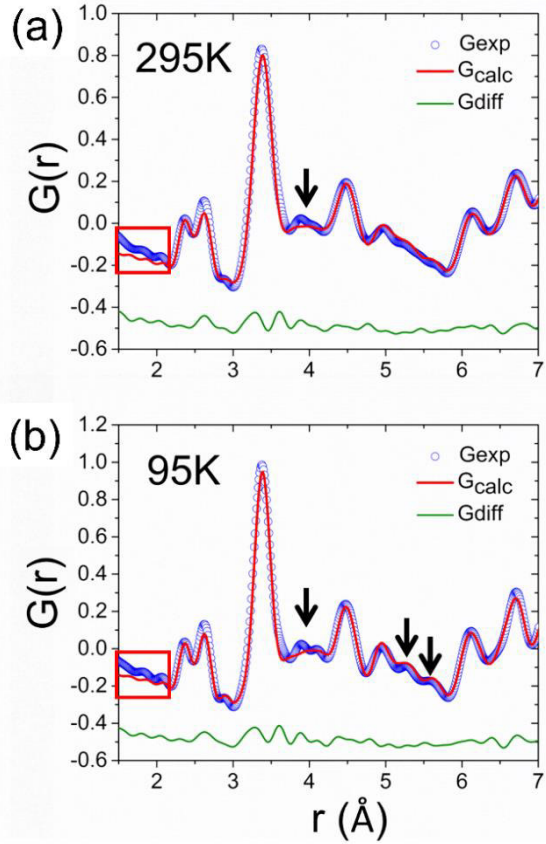

**Fig. S2:** PDF data recorded at (a) 295 K and (b) 95 K. The red boxes in both (a) and (b) show the termination ripples due to the termination of the total scattering data at a finite momentum transfer of $Q_{max}$ = 36 Å.

The effect of the increased resolution at 95 K compared to 295 K is clearly visible (*cf.* **Fig 2** and **6** in the main text). This does not mean, that these additional features are measurement artifacts. On the contrary, this makes the experiment so convincing since we were able to resolve the Sb-5s lone pair effect (which causes the distortion in the SbSe$_6$ octahedrons and the observed changes in the local Se-Se correlations of the Se atoms surrounding Sb) by cooling and by measuring up to a large momentum transfer of $Q_{max}$ = 36 Å. Both cooling and the high momentum transfer made it possible to observe the effect of the Sb-5s lone pairs on the local crystal structure in form of a clear peak splitting observed at r ≈ 4 Å, which can clearly not be described by the PDF model based on the average (i.e. long-range) crystal structure (*cf.* **Fig. 6** in the main text). Such small changes in the PDF features due to slight differences in the local and average (i.e. long-range) crystal structure can easily be missed when truncating the X-ray data at lower Q values as has convincingly been shown by Proffen et al. in 2003. **[S1]** In our case a very high resolution (large $Q_{max}$ and minimized thermal vibrations) was necessary to observe these very subtle Sb-5s lone pair dynamics. The peak splitting at r ≈ 4 Å, which is not observed at 295 K is the most convincing signature of the Sb-5s lone pair effect.

There can be 2 reasons for not observing the peak splitting at 295 K: a) the resolution is not high enough to resolve it due to thermal broadening or the Sb-5s lone pairs are spherical in average due to the atomic vibrations, i.e. the SbSe$_6$ octahedrons appear undistorted in average due to thermal vibrations of the atoms. In either case, we can safely conclude, that the differences we observe between local and average crystal structure involves Sb-5s lone pair dynamics.

3. Cell parameter evolution as a function of temperature

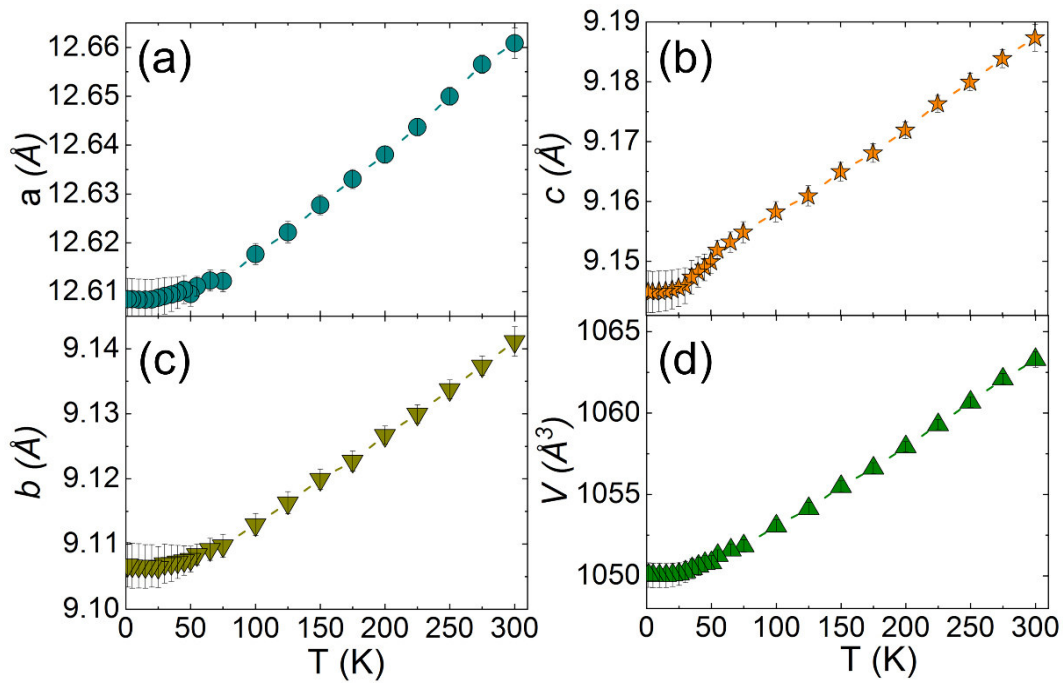

**Fig. S3:** (a) – (c): unit cell parameters and (d) cell volume evolution with temperature; all data shown in (a)-(d) were obtained from Rietveld refinements.

4. Magnetic susceptibility

The magnetic moment obtained from our magnetic susceptibility data (*cf.* **Fig. 3** and **Fig. S4**) is slightly higher (i.e. 6.8 $\mu_B$) compared to the moment of 5.3 $\mu_B$ found by Wang et al. **[S3]** as well as the magnetic moment of 4.13 $\mu_B$ (5.8 $\mu_B$ without constraint) obtained from our NPD data (*cf.* **Fig. 4** in the main text) One possible explanation is, that a very small amount of a magnetic impurity (not observed during the NPD and PDF experiments since χ(T) measurements are much more sensitive towards magnetic impurities) in the paramagnetic state has a small contribution to the magnetic moment obtained from the χ(T) curve. This hypothesis is supported by the barely visible differences between the ZFC and FC magnetic susceptibility curves shown in **Fig. S4**. The reason, that we do not observe this contribution in our NPD data is, that we only detect contributions from long-range magnetically ordered phases which are not in the paramagnetic state. Apparently the amount of possible magnetic impurities is negligible in the context of our NPD and PDF data analyses since the magnetic moment obtained from our NPD data is close to the expected value of 5 $\mu_B$ and there is no sign of any impurity phase neither in the NPD patterns nor in the PDF data.

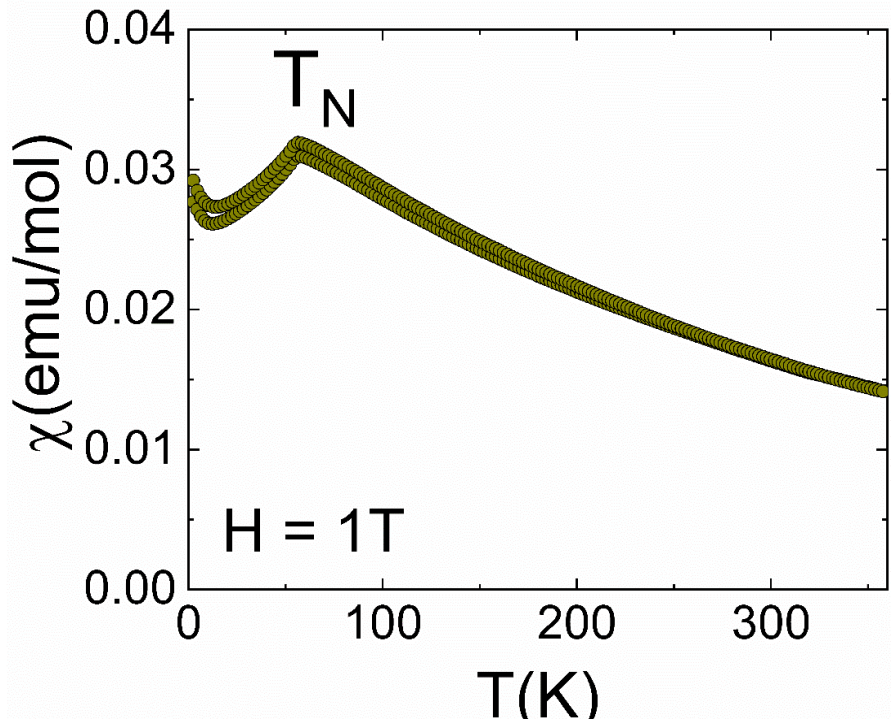

**Fig. S4:** Comparison between ZFC and FC magnetic susceptibility curves. Slight deviations at low temperatures and the rather large magnetic moment of 6.8 $\mu_B$ hint towards small amounts of magnetic impurities, which were not observed during our NPD and PDF experiments resulting in a magnetic moment of 4.13 $\mu_B$ (5.8 $\mu_B$ without constraint), which is much closer to the magnetic moment reported by Wang et al. (5.3 $\mu_B$) and the expected magnetic moment of 5 $\mu_B$ for an $Fe^{3+}$ spin state with S = 5/2.

5. Local pairwise correlations and interatomic distances in the average crystal structure – a comparison

**Table S1** summarizes all relevant local pairwise correlations and interatomic distances in the average crystal structure. While all Se-Se correlations are compared to the structural model obtained from single crystal X-ray diffraction data by Wang et al. [S3] all Fe-Fe correlations are compared to the interatomic distances obtained from our neutron diffraction data at 1.5 K. The corresponding Fe-Fe distances are labeled from J1-J5 in **Fig. 5** in the main text. Please note, that slight differences in the interatomic distances are due to thermal expansion and different experimental methods used to extract them. However, these minor differences are negligible since the pairwise correlations can still be clearly assigned (or labeled as missing) to the interatomic distances in the average crystal structure. It is important to note, that in the case of the Se-Se correlations within the $SbSe_6$ octahedrons, clear differences are observed between local and average crystal structure (*cf.* **Fig. 6** in the main text). This is not the case for the Fe-Fe correlations (*cf.* **Fig. 5** in the main text), where calculated and experimental PDF data match perfectly. We therefore conclude that Sb-5s lone pair dynamics can explain these observations with Se-mediated magnetic super exchange being a much less likely alternative.

**Table S1** Relevant pairwise correlations compared to interatomic distances in the average crystal structure

| Pairwise correlation | PDF peak position (295 K) | PDF peak position (95 K) | Interatomic distances in the average crystal structure |
|---|---|---|---|
| Se-Se | 3.887 Å | 3.884 Å | 3.841 Å [S3] |
| Se-Se | - | 4.098 Å | 4.072 Å [S3] |
| Se-Se / Ba-Ba | 4.477 Å | 4.477 Å | 4.416 Å / 4.420 Å [S3] |
| Ba-Ba | 4.977 Å | 4.961 Å | 4.950 Å [S3] |
| Se-Se | - | 5.283 Å | 5.298 Å [S3] |
| Se-Se | - | 5.598 Å | 5.898 Å [S3] |
| Fe-Fe | 6.132 Å | 6.113 Å | 6.008 Å (J1) |
| Fe-Fe | 6.713 Å | 6.699 Å | 6.496 Å (J2-2) |
| Fe-Fe | 7.249 Å | 7.241 Å | 7.482 Å (J3) |
| Fe-Fe | 8.100 Å | 8.091 Å | 8.049 Å (J4) |
| Fe-Fe | 9.285 Å | 9.241 Å | 9.105 Å (J5) |

6. RMC modelling and partial PDF analyses

This is confirmed by our RMC and partial PDF analyses (see main text), which are extended here (*cf.* **Fig. S5**). The corresponding reduced structure function F(Q) obtained from our RMC simulations is shown in **Fig S6**, while **Fig. S7** shows the RMC model for the entire experimental r-range. **Table S2** contains a summary of all RMC runs including accepted, generated and tested moves.

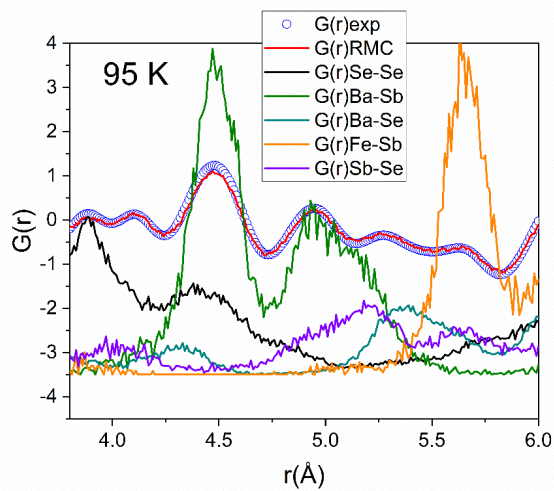

**Fig. S5:** Local pairwise correlations contributing to the PDF data in the r-range of the second coordination sphere.

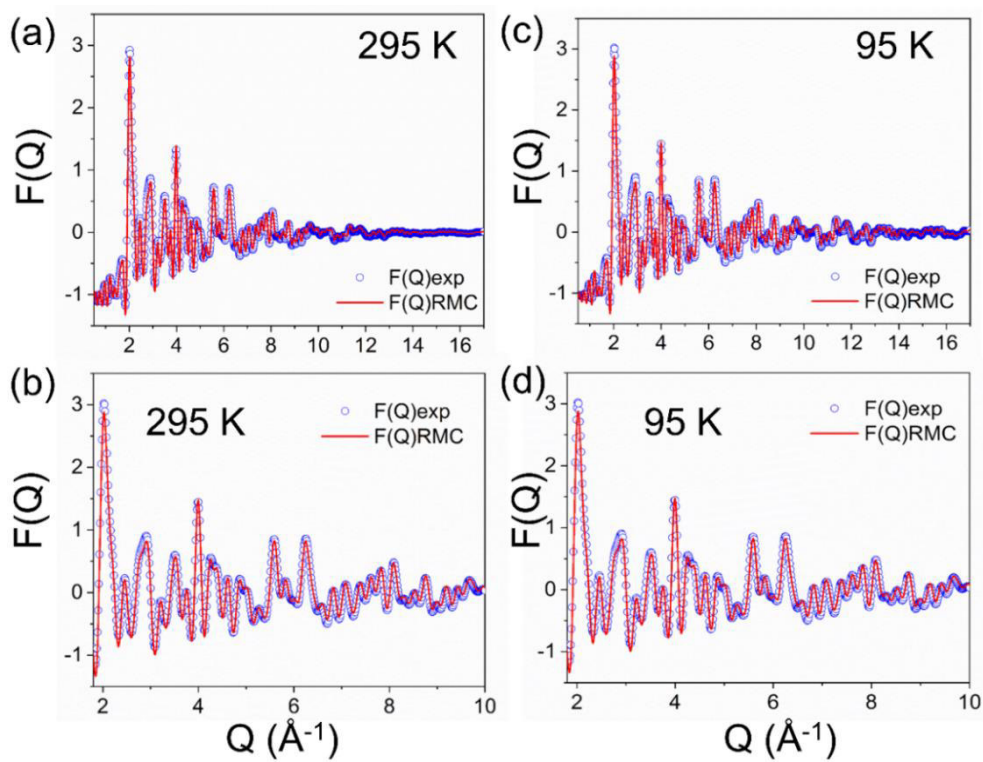

**Fig. S6** Reduced structure function F(Q) at (a) 295 K and (c) 95 K; (b) and (d) show the corresponding low-r ranges.

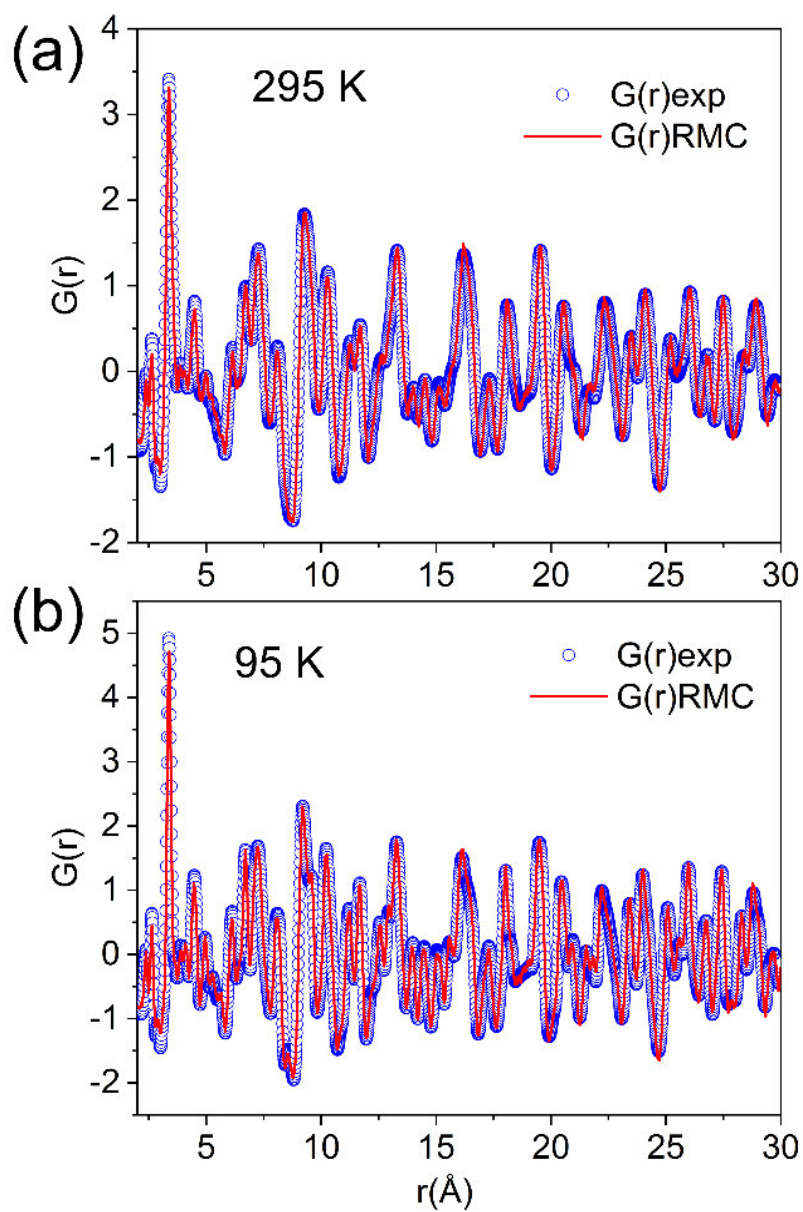

**Fig. S7** RMC model for the full experimental r-range showing excellent agreement with the experimental data for the entire experimental r-range.

**Table S2** Summary of all RMC runs

| Run # | accepted | generated | tested | $\chi^2$/DOF |
|---|---|---|---|---|
| 1 | 139206 | 3141829 | 1379269 | 104.4 |
| 2 | 296742 | 832003 | 808316 | 26.34 |
| 3 | 536116 | 1701700 | 1653374 | 21.77 |
| 4 | 437938 | 1443703 | 1403307 | 20.39 |
| 5 | 478467 | 1525201 | 1525200 | 18.48 |

7. <u>Magnetoelastic coupling – how could it look at 58 K?</u>

The differences between average and local crystal structure as well as the changes of the latter with temperature are apparently not correlated with magnetic ordering at 95 K. Magnetic PDF or $\mu^+$SR techniques can provide further insight into possible local magnetic ordering or spin fluctuations and further PDF experiments at lower temperatures will provide further insight into the nature of the magnetoelastic coupling observed previously.**[S2]** The proposed magnetic structure contains antiferromagnetic (J2-1) and ferromagnetic spin arrangements (J2-2) along the *a*-axis. Both Fe-Fe distances are, however the same (6.5 Å). Hence, magnetostructural changes might occur in order to stabilize these orderings, i.e. the Fe-Fe distances of the *J2-1* ordering become shorter than those of the *J2-2* ordering. The P$_a$2$_1$/m Shubnikov group has a monoclinic supercell ($a_m$ ~ 18.4 Å, $b_m$ ~ 9.1, $c_m$ ~ 15.6 Å and $\beta_m$ ~ 126°). In this supercell, a Γ4+ distortion mode (with a k-vector of (0,0,0) and direction *a* results in the isotropic subgroup P2$_1$/m (No. 11)). Such a distortion mode allows to move Fe atoms along the *a*-direction with two Fe rows shifting in opposite *z* directions and it can relax the *J2-1* and *J2-2* distances in order to stabilize these spin arrangements. However, such a monoclinic distortion of the crystal structure is not seen in the neutron diffraction data reported here. Further studies using low temperature synchrotron and/or single crystal diffraction techniques will most certainly be valuable for a more detailed investigation of possible magnetoelastic effects in this compound, which is beyond the scope of this manuscript.

<u>References</u>

[S1] Th. Proffen, S. J. L. Billinge, T. Egami, D. Louca, *Z. Kristallogr.* **2003**, *218*, 132.

[S2] S. Maier, C. Moussa, D. Berthebaud, F. Gascoin, A. Maignan, *Appl. Phys. Lett.* **2018**, 112, 202903.

[S3] J. Wang, J. T. Greenfield, K. Kovnir, *J. Solid State Chem.* **2016**, 242, 22.